\documentclass[aps,amsmath,showpacs,amsfonts,10pt]{revtex4}
\usepackage{epsfig,graphicx}
\usepackage[english]{babel}
\usepackage{amsfonts}
\usepackage{amsmath}
\usepackage{latexsym}
\usepackage{graphics,bm}
\usepackage{dcolumn}
\usepackage{bm}
\usepackage{rotating}
\begin{document}

\title{Luttinger liquid in superlattice structures: atomic gases, quantum dots and the classical Ising chain}

\author{Aranya B Bhattacherjee$^{1}$, Pradip Jha$^{2}$,Tarun Kumar$^{3}$ and ManMohan$^{3}$}

\address{$^{1}$Department of Physics, ARSD College, University of Delhi (South Campus), New Delhi-110021, India}\address{Department of Physics, DDU College, University of Delhi, India} \address{$^{3}$ Department of Physics and Astrophysics, University of Delhi, Delhi-110007, India}

\begin{abstract}
We study physical properties of a Luttinger liquid in a superlattice which is characterized by alternating two tunneling parameters. Employing the Bosonization approach, we describe the corresponding Hubbard model by the equivalent Tomonoga-Luttinger model. We analyze the spin-charge separation and transport property as the difference between the two tunneling parameter increases. We suggest that cold Fermi gases trapped in a bichromatic optical lattice and coupled quantum dots offer the opportunity to measure these effects in a convenient manner. We also study the classical Ising chain with two tunneling parameters. We found that the classical two-point correlator decreases as the difference between the two tunneling parameter increases.
\end{abstract}

\pacs{03.75.Lm,03.75.Kk,05.30.Jp,32.80Pj,42.50.Vk,42.50pq}

\maketitle

\section{Introduction}
Quantum many-body systems of one-dimensional interacting fermions have attracted enormous interest for more than 5 decades. Contrary to what happens in two and three dimensions, these systems cannot be described by the Landau theory of normal Fermi liquids. The appropriate paradigm for 1D interacting fermions is instead provided by the Luttinger liquid concept introduced by Haldane. The distinctive feature of the Luttinger liquid is that its low-energy excitations are collective oscillations of the charge or the spin density, as opposed to individual quasiparticles that carry both charge and spin. This leads immediately to the phenomenon of spin-charge separation, i.e the fact that the low energy spin and charge excitations of 1D interacting fermions are completely decoupled and propagate with different velocities.
Despite the firm theoretical basis upon which the Luttinger liquid theory rests, there has been precious little compelling experimental evidence that real one-dimensional electron gases are anything but Fermi liquids. In recent years, it has become possible to fabricate single channel quantum wires but unwanted impurity causes backscattering and localization, thus destroying the Luttinger liquid phase. Fortunately, there is another experimental system which is expected to exhibit Luttinger liquid behaviour, and does not suffer from complications associated with impurities-namely one-dimensional quantum Fermi gases in optical lattices.

In this context, it has now become possible to trap ultra-cold quantum gases in quasi-1D optical lattices. Much of the theoretical work has been on 1D Bose gases but more recently Recati et al \cite{recati} have studied one-dimensional quantum gases of fermionic atoms in optical lattice using Luttinger liquid approach while Polini et al. \cite{polini} have studied spin-drag and spin charge separation of atomic Fermi-Dirac gas in a one-dimensional optical lattice.
Using superposition of optical lattices with different periods \cite{Peil03}, it is now possible to generate more sophisticated periodic potentials characterized by a richer spatial modulation, the so-called optical superlattices. An important and exciting application of optical superlattice is quantum computation \cite{Sebby06}.  The physics of one-dimensional optical superlattices has been a subject of recent research, including fractional filling Mott insulator (MI) domains \cite{Bounsante04}, dark \cite{Louis04}and gap  \cite{Louis05} solitons, the Mott-Peierls transition \cite{Dimtrieva68}, non-mean field effects \cite{Rey04}, phase-diagram in two colour superlattices \cite{Roth03}, Bloch-Zener and dipole oscillations \cite{Breid07}, collective oscillations \cite{Chun05} and Bloch and Bogoluibov spectrum \cite{Bhattacherjee07}. In section II, we discuss the influence of such a kind of superlattice structure on the spin-charge separation and the compressibility of the Fermi gas.

Often referred to as artificial atoms, semiconductor quantum dots offer an unprecedented possibility of constructing at will and exploring situations ranging from practically single atom to a fully solid state many-body systems. The nanofabrication possibilities of tailoring structures to desired geometries and specifications, and controlling the number and mobility of electrons confined within a region of space, makes these structures unique tools to study transport properties. Quantum transport in arrays of tunnel coupled quantum dots have attracted attention for the past few years \cite{petrosyan,green,wege,stafford,gossard}. The controllable quantum properties of the electron in such systems opens the possibility of their application to schemes of quantum computers \cite{loss}. In section III, We study the conductivity of an electron that experiences an asymmetric tunneling when going to the left and right in an one dimensional tunnel coupled quantum dots.

The spin $1/2$ Ising chain is considered the prototypical system for quantum phase transitions \cite{sachdev}. However very little is known in literature about its classical counterpart. In section IV, we discuss the classical Ising model in the absence of an external field when the exchange energy between site $j$ and $j+1$ is not the same as $j$ and $j-1$. With the development of molecular beam epitaxy, it is now possible to envisage a superlattice in which the exchange constant varies from layer to layer. Very often one finds interesting properties in these systems \cite{weller}. Magnetic excitations in superlattices were studied in numerous works (see \cite{morkowsky} for a brief review). Hinchey and Mills \cite{hinchey} have investigated a superlattice structure with alternating ferromagnetic and antiferromagnetic layers. A common feature that connects the three systems studied in this work is the asymmetric tunneling.

\section{The Hubbard Hamiltonian and the equivalent Tomonaga-Luttinger model}

Our aim will be to study strongly correlated systems in one spatial dimension. These are typically systems of interacting electrons but we will be interested in cold Fermi gases also. The prototypical interacting electron system is the Hubbard model. This is the lattice model whose Hamiltonian in one dimension is

\begin{equation}\label{F-H}
H_{FH}=-\sum_{j,\sigma}J_{j}\left( \hat c_{j+1, \sigma}^{\dagger} \hat c_{j, \sigma}+ \hat c_{j, \sigma}^{\dagger} \hat c_{j+1, \sigma}\right)+U \sum_{j} \hat n_{j,\uparrow} \hat n_{j, \downarrow}
\end{equation},

The first term describes the hopping process, in which an electron can move from one site to the next with site dependent amplitudes $J_{j}$ which takes two distinct values, $J_{0}+(-1)^{j}\dfrac{\Delta_{0}}{2}$. The hopping process preserves the spin projection $\sigma$. For cold Fermi gases, such a kind of hopping terms can be created by superposition of two optical lattice of different periodicity (as described in section 3) while for electron system, coupled quantum wells with appropriate voltages can generate such a hopping term (as described in section 4).The second term describes the local Coulomb repulsion ($U>0$) between opposite spin electrons residing on the same site. For cold Fermi gases this would be the two body interaction as discussed in the next section. The $\hat c_{j, \sigma}$ operators are the usual annihilation operators with anti-commutation relations. Also, $\hat n_{j,\sigma}=\hat c_{j,\sigma}^{\dagger}\hat c_{j,\sigma}$ is the Fermionic number operator. In the next section, we will show that the Hamiltonian for the cold Fermi gas in an one dimensional bichromatic optical lattice can be reduced to the Fermi-Hubbard Hamiltonian \ref{F-H}.
In this section, we will derive the equivalent bosonized Hamiltonian of the Fermi-Hubbard Hamiltonian. The technique of bosonization is a powerful tool to study the spectrum of low-lying excitations and correlation functions of one-dimensional systems. Let us first look at the non-interacting limit ($U=0$). In this case, the Hamiltonian can be easily diagonalized by means of Fourier transformation. We define

\begin{equation}
c_{j,\sigma}=\dfrac{1}{\sqrt{L}}\sum_{k}\left\lbrace c_{k,\sigma}^{g}-i (-1)^{j} c_{k, \sigma}^{e}\right\rbrace e^{i2kdj}
\end{equation}

Here $L$ is the number of lattice sites with periodic boundary conditions $c_{j+L, \sigma}=c_{j, \sigma}$. As the fermions move from one well to the next, it acquires as additional phase, which depends on the height of the barrier. As the height alternates, the phase also alternates. This picture is conveniently represented by the $j$ dependent inverse Fourier transform of equation (2).
Substituting equation (2) in the non-interacting fermi-Hubbard Hamiltonian, we have

\begin{equation}
H=-2J_{0} \cos{2kd} \left\lbrace c_{k,\sigma}^{g \dagger} c_{k,\sigma}^{g}-c_{k,\sigma}^{e \dagger}c_{k,\sigma}^{e}\right\rbrace-\Delta_{0} \sin{2kd} \left\lbrace c_{k,\sigma}^{g \dagger} c_{k,\sigma}^{e}+c_{k,\sigma}^{e \dagger}c_{k,\sigma}^{g}\right\rbrace
\end{equation}

Finally, $H$ can be brought to diagonal form by defining operators

\begin{equation}
c_{k,\sigma}^{g}=\sqrt{\dfrac{1}{2}+\dfrac{J_{0} \cos{2kd}}{\epsilon_{k}}}f_{k,\sigma}= \alpha f_{k,\sigma}
\end{equation}

\begin{equation}
c_{k,\sigma}^{e}=\sqrt{\dfrac{1}{2}-\dfrac{J_{0} \cos{2kd}}{\epsilon_{k}}}f_{k,\sigma}= \beta f_{k,\sigma}
\end{equation}

Where $|\alpha_{k}|^{2}+|\beta_{k}|^{2}=1$ and $\epsilon_{k}=\sqrt{4 J_{0}^{2} \cos^{2} {2kd}+ \Delta_{0}^{2} \sin^{2} {2kd}}$.
This yields:

\begin{equation}
H=-\sum_{k,\sigma} \epsilon_{k} f_{k,\sigma}^{\dagger}f_{k,\sigma}
\end{equation}

The ground state for $N$ fermions corresponds to filling up all the states, from the lowest energy up, until the $N$ lowest-energy orbitals are filled up (taking into account spin degeneracy). The highest occupied level is the Fermi level, its energy the Fermi energy $E_{F}$ and its wave-vector the Fermi wave-vector $k_{F}$.
The relationship between $N$ and $k_{F}$ is $N=2k_{F}L/ \pi$ or $n=N/L=2k_{F}/ \pi$. When we take into account interactions and if $U<<J_{j}$ (perturbative region), it is natural to assume that only low energy states will be affected. This is reasonable within second order perturbation theory. We now introduce in the usual way the right movers (around $+ k_{F}$) and left movers (around $-k_{F}$). We then have two linearized spectrum around the two Fermi points

\begin{equation}
\epsilon(k)=v_{F}(k-k_{F})\Longrightarrow Right moving branch
\end{equation}

\begin{equation}
\epsilon(k)=-v_{F}(k+k_{F})\Longrightarrow Left moving branch
\end{equation},

according to the sign of the velocities. The relationship between this spectrum and the lattice one is given by:

\begin{equation}
v_{F}=\dfrac{\partial \epsilon(k)}{\partial k}|_{k=k_{F}}=\dfrac{d \sin{(4 k_{F}d) }(4 J_{0}^{2}-\Delta_{0}^{2})}{\hbar \sqrt{(4 J_{0}^{2} \cos^{2} {2 k_{F} d}+\Delta_{0}^{2} \sin^{2} {2 k_{F}d})}}
\end{equation}

Note that, we effectively restrict ourselves to low energies. Corresponding to the right and left moving fermions, we can introduce fermion annihilation(creation) operators $c_{\nu,\sigma}^{\dagger}$, where $\nu=R,L$ and the respective density fluctuation operators $\rho_{\nu,\sigma}(q)=\sum_{k}c_{\nu,\sigma}^{\dagger}(k+q)c_{\nu,\sigma}(k)$. Note that $\rho_{\nu,\sigma}(-q)= \rho_{\nu,\sigma}^{\dagger}(q)$. These particle-hole excitations can be written in terms of bosonic creation and annihilation operators:

\begin{equation}
\rho_{\nu,\sigma}(q)= \left \{ \begin{array} {cc}
\sqrt{\dfrac{Lq}{2 \pi}}b_{q,\sigma}^{\nu \dagger} & \mbox{$q > 0$};\\
\sqrt{\dfrac{L|q|}{2 \pi}}b_{-q,\sigma}^{\nu } & \mbox{$q < 0$} \end{array}\right .
\end{equation}

The normally ordered number operator is defined as:

\begin{equation}
\hat N_{\nu,\sigma}=\sum_{k}:c_{\nu,\sigma}^{\dagger}(k)c_{\nu,\sigma}(k):,  \nu=L,R
\end{equation}

and

\begin{equation}
\hat N_{\nu,\lambda=c,s}=\dfrac{1}{\sqrt{2}}\left( \hat N_{\nu,\uparrow} \pm \hat N_{\nu,\downarrow}\right)
\end{equation}

As a part of the bosonization process, we also introduce boson field operators:

\begin{equation}
\phi_{R,L,\sigma}=\dfrac{i}{\sqrt{L}}\sum_{q>0}\dfrac{1}{\sqrt{q}}e^{-\alpha q/2}\left( e^{\pm iq x}b_{q,\sigma}^{R,L}-e^{\mp iq x} b_{q,\sigma}^{R,L \dagger}\right)
\end{equation}

\begin{equation}
\phi_{\nu,\lambda=c,s}=\dfrac{1}{\sqrt{2}}\left(\phi_{\nu,\uparrow}(x) \pm  \phi_{\nu, \downarrow}(x)\right)
\end{equation}

\begin{equation}
b_{q,\lambda=c,s}^{\nu}=\dfrac{1}{\sqrt{2}}\left(b_{q,\uparrow}^{\nu} \pm b_{q,\downarrow}^{\nu} \right)
\end{equation}

The bosonized Hamiltonian is written as:

\begin{equation}
H=H_{0}+H_{int}
\end{equation}

\begin{equation}
H_{0}=\hbar \sum_{\lambda}\left\lbrace \dfrac{v_{F}}{2} \int_{-L/2}^{L/2} dx \sum_{\nu} :\left( \partial_{x} \phi_{\nu \lambda}\right)^{2}+\dfrac{\pi v_{F}}{L} \sum_{\nu}\hat N_{\nu \lambda}^{2} :\right\rbrace
\end{equation}

\begin{equation}
H_{int}= \hbar \sum_{\lambda} \left\lbrace \dfrac{1}{L}\left[ \dfrac{g_{4 \lambda}}{2}\left( \hat N_{R \lambda}^{2}+\hat N_{L \lambda}^{2}\right)+g_{2 \lambda} \hat N_{R \lambda}\hat N_{L \lambda} \right] \right\rbrace+\hbar \sum_{\lambda} \int_{-L/2}^{L/2}\dfrac{dx}{2 \pi}\left\lbrace \dfrac{g_{4 \lambda}}{2} \sum_{\nu} :\left( \partial_{x} \phi_{\nu \lambda}\right)^{2} : -g_{2 \lambda}:\left(\partial_{x} \phi_{R \lambda} \right)\left( \partial_{x} \phi_{L \lambda}\right)  :\right\rbrace
\end{equation}

Where, $g_{i, \lambda=c,s}=g_{i,\parallel} \pm g_{i, \perp}$, $i=2,4$. $g_{i,\parallel}$ is the intraspecies interaction which is zero for cold Fermionic atoms. $g_{i, \perp}=U d$ is the interspecies interaction for cold Fermionic atoms. $g_{2 \lambda}$ is the strength of the forward scattering between particles belonging to different branches and with different spin state, while $g_{4 \lambda}$ is the strength of the forward scattering between particles belonging to same branches and with different spin state. We now have two decoupled sectors corresponding to charge and spin excitations. The Hamiltonian in equation (), can be diagonalized by the following Bogoliubov transformation:

\begin{equation}
\hat d_{q,\lambda}^{1}= \cosh{\gamma} \hat b_{q, \lambda}^{R}+\sinh{\gamma} \hat b_{q,\lambda}^{L \dagger}
\end{equation}

\begin{equation}
\hat d_{q,\lambda}^{2}= \sinh{\gamma} \hat b_{q, \lambda}^{R}+\cosh{\gamma} \hat b_{q,\lambda}^{L \dagger}
\end{equation}

Where, $\tanh{2 \gamma}=\Lambda=\dfrac{g_{2 \lambda}}{2 \pi v_{F}+g_{4 \lambda}}$. Finally, the diagonalized Hamiltonian is:

\begin{equation}\label{diag Hamil}
H=\hbar \sum_{\mu \lambda} u_{\lambda} \sum_{q>0}q \hat d_{q \lambda}^{\mu \dagger} \hat d_{q \lambda}^{\mu} +\dfrac{\hbar \pi}{2 L}\left( v_{N \lambda}\hat N_{\lambda}^{2}+v_{J \lambda} \hat J_{\lambda}^{2}\right), \mu=1,2
\end{equation}

Where,

\begin{equation}
\hat N_{\lambda}=\hat N_{R \lambda} + \hat N_{L \lambda},
\end{equation}

\begin{equation}
\hat J_{\lambda}=\hat N_{R \lambda}-\hat N_{L \lambda}
\end{equation}

\begin{equation}
v_{N \lambda}=\dfrac{u_{\lambda}}{g_{\lambda}}
\end{equation}

\begin{equation}
v_{J \lambda}=u_{\lambda}g_{\lambda}
\end{equation}

\begin{equation}
g_{\lambda}=\sqrt{\dfrac{2 \pi v_{F}+g_{4 \lambda}-g_{2 \lambda}}{2 \pi v_{F}+g_{4 \lambda}+g_{2 \lambda}}}
\end{equation}

\begin{equation}
u_{\lambda}=v_{F}\sqrt{\left( 1+\dfrac{g_{4 \lambda}}{2 \pi v_{F}}\right)^{2}-\left(\dfrac{g_{2 \lambda}}{2 \pi v_{F}} \right)^{2}  }
\end{equation}

In the above discussions we have neglected the back scattering and Umklapp terms. For the case of cold Fermi gases, this is justified if the optical lattice depth is not large. We now consider in the next two sections two specific systems namely cold Fermionic gases in optical superlattice and a system of coupled quantum dots and study some properties as a function of $\Delta_{0}/2J_{0}$.

\section{The Hubbard model for a cold Fermi gas in a bichromatic optical lattice}

\begin{figure}[t]
\hspace{-0.0cm}
\includegraphics [scale=0.5]{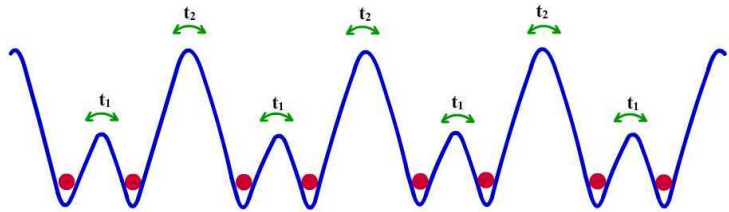}
\caption{Schematic drawing of the optical superlattice with alternating big and small wells.}
\label{1}
\end{figure}

We consider an elongated cigar shaped (quasi-1D) dilute gas of fermionic atoms of mass $m$ with two internal ground state, $|\sigma=\uparrow,\downarrow>$, representing a spin-$1/2$ system. We will assume that the two internal levels are equally populated i.e $N_{\uparrow}=N_{\downarrow}$. The atoms are cooled below the Fermi-degeneracy temperature $k_{B}T_{F}\sim N \hbar \omega_{l}$, $N=N_{\uparrow}+ N_{\downarrow}$ is the total number of particles. The condition for a quasi-1D system is a tight transverse harmonic trapping with frequency $\omega_{\perp}$ exceeding the characteristic energy scale of the longitudinal motion. In this way the transverse degrees of freedom are frozen. Because of quantum degenaracy, the longitudinal motion has all the energy levels up to the Fermi energy $\epsilon_{F}\sim k_{B}T_{F}$ filled. Typical values of $\omega_{\perp}$ and $\omega_{l}$ (frequency of longitudinal confinement) are in the range $2 \pi (300-400)$ Hz and $2 \pi (2-10)$ hz respectively. Thus we require the total number of particles to be restricted by $N<\omega_{\perp}/\omega_{l}$, which is typically of the order of few hundred. Because of the Pauli principle, at low temperature, only s-wave collisions between atoms in different internal states are allowed. Therefore, all the relevant interactions are characterized by inter component scattering length $a$. The strength of the effective 1D interaction is $g=\dfrac{2 \pi \hbar^{2} a}{m l_{\perp}^{2}}$, where $l_{\perp}=\sqrt{\dfrac{\hbar}{m \omega_{\perp}}}$ is the harmonic oscillator transverse length and $a<l_{\perp}$.

\begin{figure}[t]
\hspace{-2.0cm}
\includegraphics{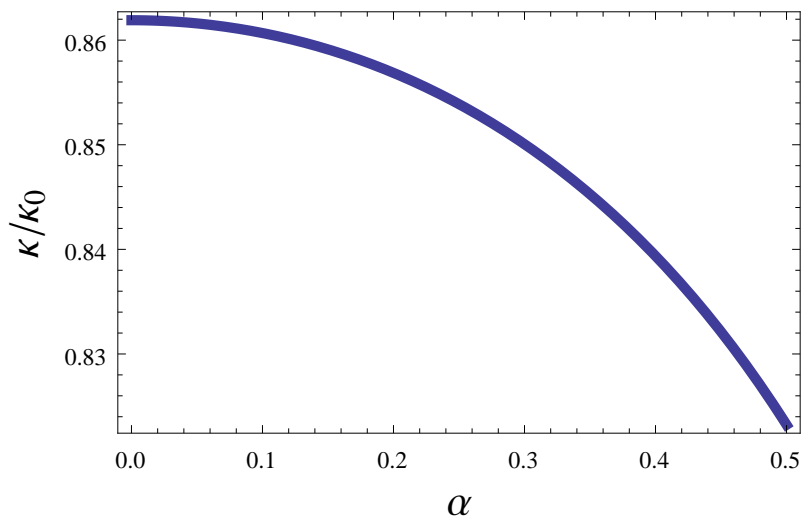}
\caption{Plot of normalized compressibility as a function of $\alpha$ for $U/2J_{0}=0.2$ and $K_{F}d=0.1$}.
\label{2}
\end{figure}

\begin{figure}[t]

\begin{tabular}{cc}
×\includegraphics [scale=0.7]{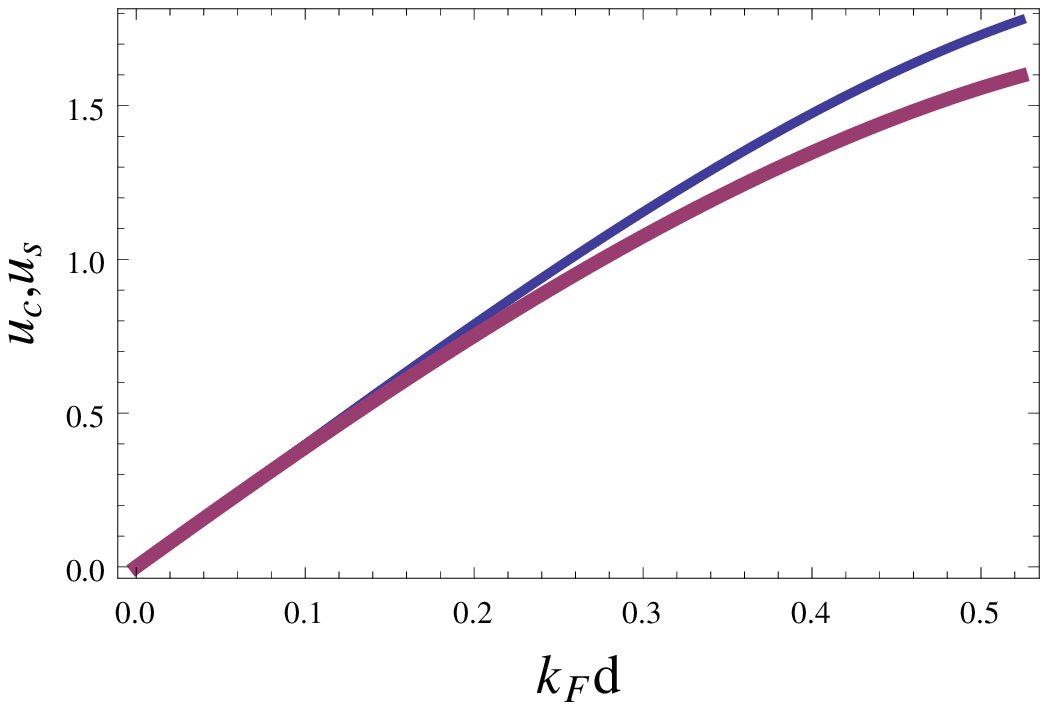}& \includegraphics [scale=0.7] {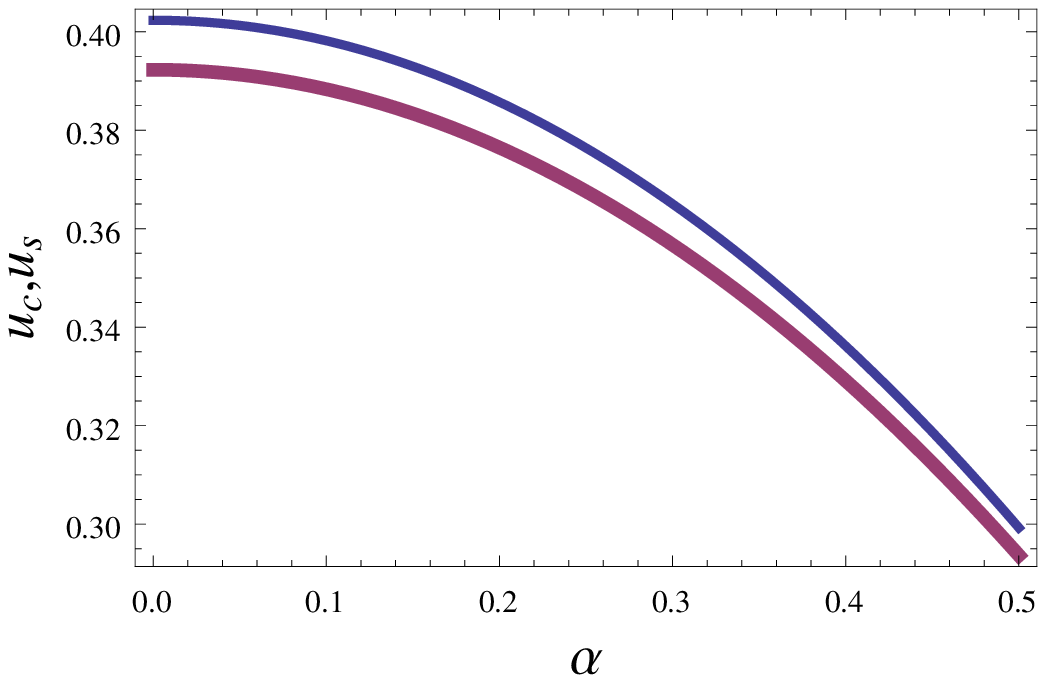}\\

\end{tabular}

\caption{Plot of the charge velocity $u_{c}$ (thin curve) and spin velocity $u_{s}$ (thick curve) as a function of $k_{F} d$ (left plot) for $\alpha=0.1$ and $U/2J_{0}=0.2$ and as a function of $\alpha$ (right plot) for $k_{F}d=0.1$. }
\label{3}
\end{figure}

Thus the system is described by the following one-dimensional Hamiltonian

\begin{equation}
H=\sum_{\sigma} \int dx \psi_{\sigma}^{\dagger}(x) \left( \dfrac{-\hbar^{2}}{2 m} \partial_{x}^{2}+V_{ext}(x)\right)\psi_{\sigma}+g \int dx \psi_{\uparrow}^{\dagger}(x) \psi_{\downarrow}^{\dagger}(x)\psi_{\uparrow}(x)\psi_{\downarrow}(x)
\end{equation},

Where $\psi_{\sigma}$ is the 1D field operators for atoms in state $\sigma$. The external potential $V_{ext}(x)=V_{L}(x)+V_{op}(x)$, includes both the longitudinal confinement $V_{L}(x)=\dfrac{1}{2}m \omega_{l}^{2} x^{2}$ and the two-colour optical lattice potential $V_{op}(x)=V_{1} \cos^{2}{\dfrac{\pi z}{d_{1}}}+V_{2}\cos^{2}{\dfrac{\pi z}{d_{2}}}$. Here $d_{1}$ and $d_{2}>d_{1}$ are respectively, the primary and secondary lattice constants. $V_{1}$ and $v_{2}$ are the respective amplitudes. The secondary lattice acts as a perturbation and hence $V_{1}>V_{2}$. We will take the particular case $d_{2}=2 d_{1}=2 d$. In addition, we will consider the case when the optical lattice dominates over the harmonic potential. We expand the atomic field operators in the lowest-band Wannier basis

\begin{equation}
\psi_{\sigma}=\sum_{j} W(x-x_{j}) \hat c_{j,\sigma},
\end{equation}

Where, $W(x-x_{j})$ is the Wannier function centered at the $j^{th}$ site and $\hat c_{j,\sigma}$ is the annihilation operator for a fermion in the $j^{th}$ site with spin $\sigma$. Sunstituting equation (2) into equation (1) and retaining only the nearest neighbour terms, we get the Fermi-Hubbard Hamiltonian:

\begin{equation}
H_{FH}=-\sum_{j,\sigma}J_{j}\left( \hat c_{j+1, \sigma}^{\dagger} \hat c_{j, \sigma}+ \hat c_{j, \sigma}^{\dagger} \hat c_{j+1, \sigma}\right)+U \sum_{j} \hat n_{j,\uparrow} \hat n_{j, \downarrow}
\end{equation},

Here, the onsite energies are taken to be zero. $J_{j}$ is the site dependent tunneling and takes two distinct values, $J_{0}+(-1)^{j}\dfrac{\Delta_{0}}{2}$. The strength of the effective on-site interaction energy is $U=g \int dx |W(x)|^4$.

The simple form of the Hamiltonian \ref{diag Hamil} makes the calculation of some physical properties rather straightforward. One important quantity, the compressibility of the gas is written as:

\begin{equation}
\dfrac{\kappa}{\kappa_{0}}=\dfrac{v_{F}g_{c}}{u_{c}}
\end{equation}

Here, $\kappa_{0}$ is the compressibility of the noninteracting gas. The density of the gas is approximated as  homogeneous. This is true if the trapping potential along the optical lattice axis is very weak and the also the depth of the optical lattice is small.
The systems interacting with replusive interactions, the optical lattice reduces the compressibility of the system, since the effect of repulsion is enhanced by the squeezing of the condensate wave function in each well. $u_{c}/g_{c}$ fixes the energy needed to change the particle density. From figure 2, we note that the compressibility decreases with the parameter $\alpha$. This means that with increasing strength of the secondary lattice, the energy required to change the particle density increases. Figure 3 shows a plot of the spin velocity ($u_{s}$) and charge velocity($u_{c}$) as a function of the quasimomentum ($k_{F}d$) and the parameter $\alpha$. As the condensate moves across the Brillioun zone, the difference between the spin and charge velocity increases and is maximum at $\pi/4$. This suggests an effective mechanism to observe the spin-charge separation. One can move the condensate across the Brillouin zone by accelerating the condensate in the optical lattice. On the other hand the spin-charge separation decreases with increasing $\alpha$. This is perhaps due to the fact that as the strength of the secondary lattice increases, the condensate becomes more localized.

\section{Linear array of tunnel coupled quantum dots}

\begin{figure}[t]
\hspace{-0.0cm}
\includegraphics [scale=0.07]{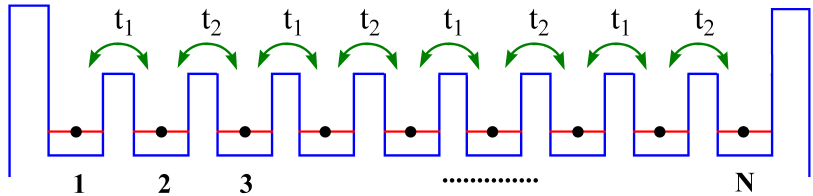}
\caption{Schematic drawing of the chain of tunnel-coupled quantum dots. Notice the alternating tunneling coefficients.}
\label{4}
\end{figure}

\begin{figure}[t]
\hspace{-0.0cm}
\includegraphics [scale=0.7]{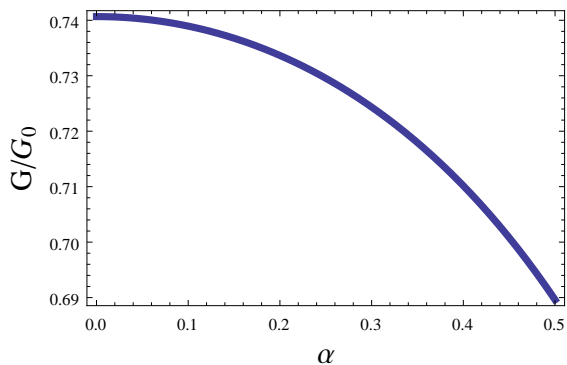}
\caption{Normalized conductance versus $\alpha$ for $U/2J_{0}=0.2$. }
\label{5}
\end{figure}

We consider electron transport in a linear array of nearly identical quantum dots (QDs) which are electrostatically defined in a two-dimensional electron gas by means of metallic gates on top of a semiconductor heterostructure (GaAs/AlGaAs). The individual tunneling rates are determined by the voltages applied to the gates defining the corresponding interdot tunneling barriers.  Thus appropriate gate voltages have to be applied to reproduce the Hamiltonian \ref{F-H}.

The conductance $G$ is found as

\begin{equation}
G=G_{0}/\sqrt{1+\gamma}
\end{equation}

Where, $G_{0}=2e^2/h$ and $\gamma=2U/(\pi \hbar v_{F})$. In figure 3, we plot the dimensionless conductivity ($G/G_{0}$) for repulsive interaction versus the parameter $\alpha=\Delta_{0}/2J_{0}$. We notice that as $\alpha$ increases, the conductivity decreases which again is an indication of localization of the electrons.
The correlation functions give information about the tendency of the system to show long range order. The correlations that decay the slowest are the dominant ones. The bosonization method makes the calculation of the correlation functions straight forward. Here we will focus on three most dominant correlations, i.e., oscillatory part of the spin density (spin density wave, $D^{z}_{sdw}\sim 1/x^{(g_{c}+g_{s})}$, $D^{x,y}_{sdw}\sim 1/x^{(g_{c}+1/g_{s})}$) and oscillatory part of the charge (charge density wave, $D_{cdw} \sim 1/x^{(g_{c}+g_{s})}$). An analysis of the factors $g_{c}$ and $g_{s}$ reveals that as $\alpha$ increases the only correlation that becomes dominant is $D^{x,y}_{sdw}$, i.e the transverse component of the spin density wave.

\section{The classical Ising chain}

It is known that the quantum transitions in the quantum Ising model in $d$ dimension is intimately connected to finite temperature phase transitions in classical Ising model in $D=d+1$ dimension. The $D=1$ and $N=1$ classical Ising model does not show any phase transition but it has regions where the correlation length becomes very large and the properties of these regions are very similar to those in the vicinity of the phase transition points in higher dimensions.  Here we will consider the $D=1$ and $N=1$ classical spin ferromagnet, more commonly known as the ferromagnetic Ising chain. This chain has the partition function

\begin{equation}
Z=\sum_{\sigma_{i}^{z}=\pm 1}exp{(-H)},
\end{equation}

where $\sigma_{i}^{z}$ are Ising spins on sites $i$ of a chain, which take the values $\pm 1$, and $H$ is given by

\begin{equation}
H=-\sum_{i=1}^{M}(J_{0}-(-1)^{i} \Delta_{0}/{2}) \sigma_{i}^{z} \sigma_{i+1}^{z}-h \sum_{i=1}^{M}\sigma_{i}^{z}
\end{equation}

Here $M$ is the total number of Ising spins and $h$ is the external magnetic field. We will assume periodic boundary conditions, therefore $\sigma_{M+1}^{z}=\sigma_{i}^{z}$. Now following Ising, we write $Z$ as a trace over a matrix product with one matrix for every site.

\begin{equation}
Z=\sum_{\sigma_{i}^{z}}\prod_{i=1}^{M} T_{1,i}(\sigma_{i}^{z},\sigma_{i+1}^{z}) T_{2}(\sigma_{i}^{z}),
\end{equation}

where

\begin{equation}
T_{1,i}(\sigma_{i}^{z},\sigma_{i+1}^{z})= \left ( \begin{array}{cc}
 {e^{(J_{0}-(-1)^i\Delta_{0}/2)}} ,   {e^{-(J_{0}-(-1)^i\Delta_{0}/2)}} \\ {e^{-(J_{0}-(-1)^i\Delta_{0}/2)}} , {e^{(J_{0}-(-1)^i\Delta_{0}/2)}}
\end{array}  \right )
\end{equation}

and

\begin{equation}
T_{2}(\sigma_{i}^{z})=\left ( \begin{array}{cc}
e^{h}, 0 \\ 0, e^{-h}
\end{array} \right )
\end{equation}

$T_{1,i}$ will have values different for $i$ even and $i$ odd. The matrix $T_{1} T_{2}$ is identified as the transfer matrix of the Ising chain. Let us now define a matrix $T_{3}=T_{1,i=even} T^{-1}_{1,i=odd}$. In the limit $M\rightarrow \infty$ one can show that

\begin{equation}
Z=Tr [T_{1}^{2} T_{2}^{2} T_{3}]^{M/2}=\epsilon_{1}^{M/2}+\epsilon_{2}^{M/2},
\end{equation}

where $\epsilon_{1}$ and $\epsilon_{2}$ are the eigenvalues of $T_{1}^{2} T_{2}^{2} T_{3}$. For the case $h=0$ (no magnetic field), $T_{2}=1$. The eigenvalues are found as

\begin{figure}[t]
\hspace{-0.0cm}
\includegraphics [scale=0.7]{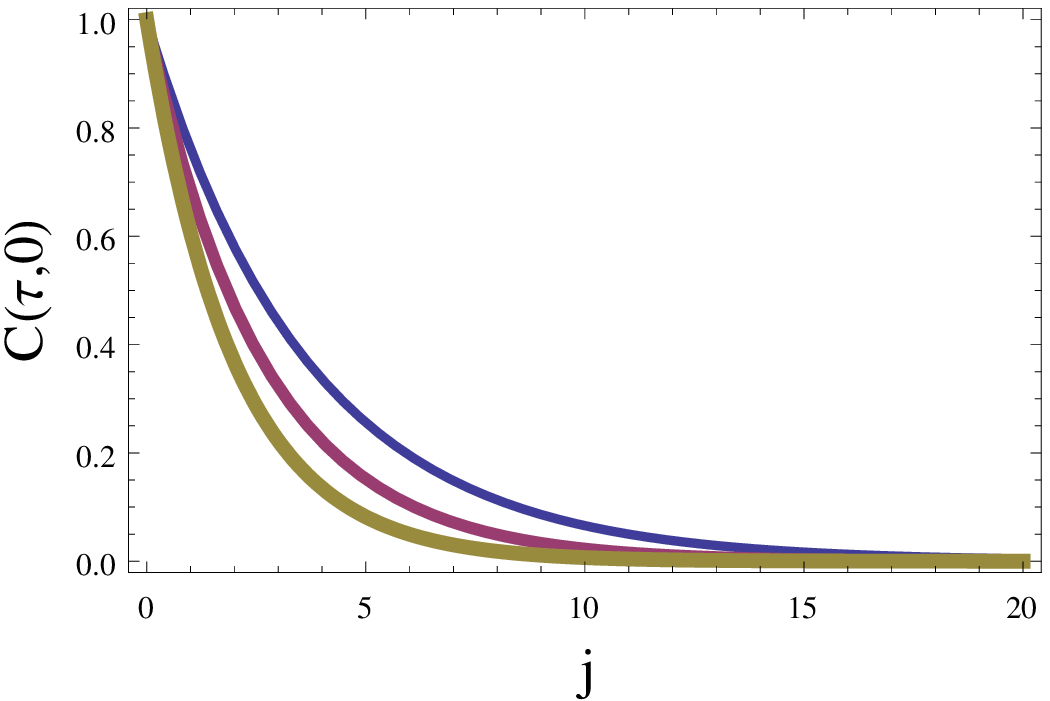}
\caption{Two point correlator as a function of lattice site $j$ for three values of $\Delta_{0}/2J_{0}=0.01, 0.25, 0.5$}. The thickness of the plots increases as the value of $\Delta_{0}/2J_{0}$ increases.
\label{6}
\end{figure}

\begin{equation}
\epsilon_{1}=\dfrac {2cosh (J_{0})}{\sqrt{sinh(2 J_{0}+\Delta_{0})}} \sqrt{sinh(2 J_{0})+sinh(\Delta_{0})}
\end{equation}

\begin{equation}
\epsilon_{2}=\dfrac {2sinh (J_{0})}{\sqrt{sinh(2 J_{0}+\Delta_{0})}} \sqrt{sinh(2 J_{0})-sinh(\Delta_{0})}
\end{equation}

The eigenvalue $\epsilon_{1}$ increases with increasing $\Delta_{0}$ and saturates at $1.2$ at $\delta_{0}=2 J_{0}$ while the other eigenvalue $\epsilon_{2}$ goes to zero at $\Delta_{0}=2 J_{0}$.

Now we calculate the correlation function exactly. For simplicity we consider the case of zero external field ($h=0$) and describe the two-point correlator

\begin{equation}
<\sigma_{i}^{z} \sigma_{j}^{z}>=\dfrac{1}{Z} \sum_{\sigma_{i}^{z}} e^{(-H)} \sigma_{i}^{z} \sigma_{j}^{z}.
\end{equation}

In the limit of an infinite chain, the two-point correlator in terms of continuous variables is derived as

\begin{equation}
C(\tau,0)= <\sigma(\tau) \sigma(0)> e^{-|\tau|/\xi},
\end{equation}

where the correlation length $\xi$ is written as,

\begin{equation}
\dfrac{1}{\xi}=\dfrac{1}{a} \left \{  ln [coth (J_{0})]+\dfrac{1}{2} ln \left [\dfrac{sinh(2 J_{0})+sinh(\Delta_{0})}{sinh(2 J_{0})-sinh(\Delta_{0})} \right ] \right \}
\end{equation}

Here, $\tau=j a$, $a$ is the lattice spacing. Figure 6. displays the two point correlator as a function of lattice site $j$ for three values of $\Delta_{0}/2J_{0}=0.01, 0.25, 0.5$. Clearly, we see that as $\Delta_{0}/2J_{0}$ increases the correlation decays faster indicating the fact that the spins are getting localized.

\section{Conclusions}

In conclusion we have studied three one dimensional superlattice systems (characterized by two tunneling parameters) namely, atomic gases in one dimensional superlattice, linear one dimensional array of quantum dots and the one dimensional classical Ising chain. In particular for atomic gases, we found that as the difference between the two tunneling parameter increases, the difference between the spin and charge velocities decreases. This is attributed to the increasing localization of the atoms in the wells of the optical lattice. On the other hand for the case of linear array of quantum dots, as the difference of the two tunneling parameters increases the conductance decreases attributed to the pinning of the electrons. For the classical Ising chain, the two-point correlator decreases with increasing strength of the superlattice which is attributed to the localization of the spins. In general we conclude that atoms, electrons and spins are comparatively more localized in a superlattice structure.  This study demonstrates that by tuning the two tunneling parameters, one can coherently control the transport properties of a superlattice structure.


\begin{thebibliography}{99}

%
\bibitem{recati}
A. Recati et al., Phys. Rev. Letts., {\textbf{90}}, 020401 (2003).
%
\bibitem{polini}
M. Polini and G. Vignale,  Phys. Rev. Letts., {\textbf{98}}, 266403 (2007).
%
\bibitem{Peil03}
S. Peil et. al. Phys. Rev. A {\textbf{67}}, 051603 (R) (2003).
%
\bibitem{Sebby06}
J. Sebby-Strabley et. al. Phys. Rev A, {\textbf{73}}, 033605 (2006).
%
\bibitem{Bounsante04}
P. Buonsante and A. Vezzani, Phys. Rev. A, {\textbf{70}}, 033608 (2004); P. Buonsante, V. Penna and A. Vezzani, Phys. Rev. A, {\textbf{70}}, 061603 (R), (2004); P. Buonsante, V. Penna and A. Vezzani, Phys. Rev. A, {\textbf{72}}, 013614 (2005); P. Buonsante, V. Penna and A. Vezzani, Laser Physics, {\textbf{15}}(2), 361 (2005).

%
\bibitem{Louis04}
P. J. Y. Louis, E. A. Ostrovskaya and Y. S. Kivshar, J. Opt. B, {\textbf{6}}, S309 (2004).
%
\bibitem{Louis05}
P. J. Y. Louis, E. A. Ostrovskaya and Y. S. Kivshar, Phys. Rev. A, {\textbf{71}}, 023612 (2005), M .A. Porter, P. G. Kevrekidis, R. Carretero-Gonzalez and D. J. Frantzeskakis, Cond-mat/0507676.
%
\bibitem{Dimtrieva68}
 L. A. Dmitrieva and Y. A. Kuperin, Cond-mat/0311468.
%
\bibitem{Rey04}
A. M. Rey, B. L. Hu, E. Calzetta, A. Roura and C. W. Clark, Phys. Rev. A, {\textbf{69}}, 033610 (2004).
%
\bibitem{Roth03}
R. Roth and K. Burnett, Phys. Rev. A, {\textbf{68}}, 023604 (2003).
%
\bibitem{Breid07}
B.M. Breid, D. Witthaut and H.J. Korsch, New Jour. Phys. {\textbf{9}}, 62 (2007), B.M. Breid, D. Witthaut and H.J. Korsch, New Jour. Phys.{\textbf{8}}, 110 (2006), A. Bhattacherjee and M. Pietrzyk, Cond-mat/0701364, L. Sanchez-Palencia and L. Santos, Phys. Rev. A, {\textbf{72}}, 053607 (2005).
%
\bibitem{Chun05}
Chou-Chun Huang and Wen-Chin Wu, Phys. Rev. A, {\textbf{72}}, 065601 (2005).
%
\bibitem{Bhattacherjee07}
A. Bhattacherjee, J. Phys.B. At. Mol. Opt. Phys. {\textbf{40}}, 143 (2007).
%
\bibitem{petrosyan}
D. Petrosyan and P. Lambropoulos, Optics Communications, {\textbf{264}}, 419 (2006).
%
\bibitem{green}
A. D. Greentree et al., Phys. Rev. B., {\textbf{70}}, 235317 {2004}.
%
\bibitem{wege}
M. R. wegewijs, Y. V. Nazarov, Phys. Rev. B., {\textbf{60}}, 14318 {1999}.
%
\bibitem{stafford}
C. A. Stafford et al., Phys. Rev. B., {\textbf{58}}, 7091 {1998}.
%
\bibitem{gossard}
D. S Duncan et al., Phys. Rev. B., {\textbf{63}}, 045311, {2001}.
%
\bibitem{loss}
D. Loss and D. P. DiVincenzo, Phys. Rev. A {\textbf{57}} 120 {1998}; P. Zanardi, F. Rossi, Phys. Rev. Lett. {\textbf{81}}, 4752 {1998}; M. Friesen et al., Phys. Rev. B {\textbf{67}}, 121301(R) {2003}.
%
\bibitem{sachdev}
S. Sachdev, Quantum Phase Transitions, Cambridge University Press, Cambridge, UK, 1999.
%
\bibitem{weller}
D. Weller et al., Phys. Rev. Lett., {\textbf{54}}, 1555, {1985}, C. Rau and S. Eicher, Phys. Rev. Lett., {\textbf{47}}, 439, {1981}; C. Rau, C. Jin and M. Robert, J. Appl. Phys., {\textbf{63}}, 3667 {1988}.
%
\bibitem{morkowsky}
J. A. Morkowsky and A. Sza jek, J. Magn. Magn. Mater., {\textbf{71}}, 299 {1988}.
%
\bibitem{hinchey}
L. L. Hinchey and D. L. Mills, Phys. Rev. B, {\textbf{33}}, 3329 {1986}.
%


\end{thebibliography}
\end{document}